\documentclass{elsart}

\usepackage{amssymb}
\usepackage{amsmath}
\usepackage{epsfig}

\usepackage{color}
\newcommand{\PGRcomm}[1]{{\color{black} #1}}

\begin{document}

\begin{frontmatter}



\title{Self-interaction correction in a simple model}


\author{P.~M.~Dinh\corauthref{cor}$^a$}
\author{, J.~Messud$^a$, P.-G.~Reinhard$^b$, and E.~Suraud$^a$}

\corauth[cor]{Corresponding author\\{\it Email-address}~:
  dinh@irsamc.ups-tlse.fr} 
\address{$^a$Laboratoire de Physique Th\'eorique, Universit\'e Paul
  Sabatier, CNRS\\
  118 route de Narbonne F-31062 Toulouse C\'edex, France}
\address{$^b$Institut f{\"u}r Theoretische Physik, Universit{\"a}t
  Erlangen,\\
  Staudtstrasse 7 D-91058 Erlangen, Germany}

\begin{abstract}
We discuss various ways to handle self-interaction corrections (SIC) 
to  Density Functional Theory (DFT) calculations. To that end, we
use a simple model of few particles in a finite number of states
together with a simple zero-range interaction for which full
Hartree-Fock can easily be computed as a benchmark.
The model allows to shed some light on the balance between
orthonormality of the involved states and energy variance.
\\
\end{abstract}

\end{frontmatter}

\section{Introduction}
\label{sec:intro}

Density Functional Theory
(DFT)~\cite{Hoh64,Koh65,Par89,Dre90,Koh99,Jon89,Run84} is a standard
theoretical tool for the description of electronic systems,
which takes into account exchange and correlation
effects. Practically, DFT methods require approximations to the
exchange and correlation potentials. The most widely used is the Local
Density Approximation (LDA) \cite{Koh99}. This scheme however 
contains a spurious self-interaction. As a consequence, the Coulomb
asymptotics, the ionization potential, and the potential energy
surface of a system turn out to be wrong. These incorrect behaviors
can lead to misleading results especially in time-dependent processes, as
e.g. dynamics of ionization.

A corrected description includes a self-interaction correction (SIC)
\cite{Per81,Kum}. SIC methods were tried and tested in various domains
of physics, such as atomic, molecular, cluster and solid state
physics, see
e.g. \cite{Ped84,Ped85,Goe97,Polo,Vydrov,Ton97,Ulr00,Per79,Zun80}. The
original SIC scheme leads to an orbital dependent (and thus non
hermitian) mean-field which, as a consequence, leads to violation of
orthonormality. 
\PGRcomm{
That problem has been attacked with various strategies.
When maintaining orbital-dependent potentials, the most consistent
technique is to deal with a matrix of Lagrangian multipliers taking
care explicitely for orthonormality, see e.g. \cite{Goe97}.
A formally elegant alternative is to enforce a common mean-field
potential by the method of optimized effective potentials (OEP)
\cite{Tal76} which, however, can become technically very involved.
Thus one often steps down to the Krieger-Li-Iafrate (KLI) approximation
for OEP \cite{Kri92}. KLI-SIC is widely accepted as a useful and
inexpensive approach to SIC. There are, however, some drawbacks
showing up in critical applications. Indeed KLI is underestimating the
often necessary localization of wavefunctions \cite{Gar00}
which, e.g., leads to problems with the polarizibility in chain
molecules \cite{Gri01,Gru02} or with NMR shieldings \cite{Tea05}.
Time dependent KLI also runs in serious problems with energy
conservation and zero-force theorem \cite{Mun07}.
Thus there remains a need for a direct handling of SIC to deal with
such critical applications. The non-hermicitiy of the
orbital-dependent mean-field and proper handling of orthonormality of
the occupied single-particle states are then crucial topics to be
considered, particularly in time-dependent applications.
}
The formally
sound, but practically cumbersome, way to deal with that is to use a
full matrix of Lagrange multipliers. It is widely used practice to
abbreviate that by using standard diagonalization schemes together
with explicit orthonormalization. 
It is the aim of the present paper to compare various solution
strategies and to investigate in detail
the interplay of orthonormality and energy diagonality (or variance,
respectively). This will be done in a simplemost model involving two
active states.

The paper is organized as follows~: Section \ref{sec:mod} introduces
the model, section \ref{sec:MF} summarizes briefly the various
approaches, and section \ref{sec:compar} is devoted to discussion of
results.


\section{The two-state model}
\label{sec:mod}

\subsection{The model Hamiltonian}
\label{sec:model}

\begin{figure}[htbp]
\begin{center}
\epsfig{file=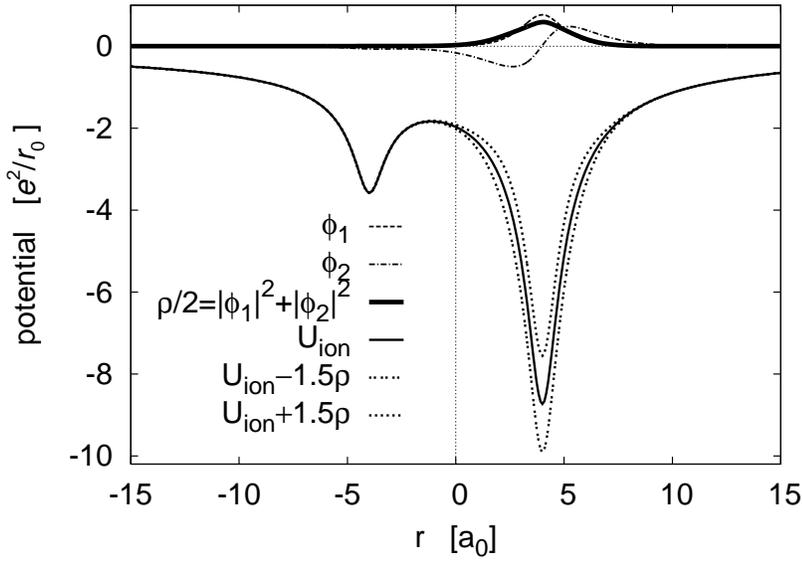,width=0.8\linewidth}
\caption{Potential $U_{\rm ion}$ (full thin line), eigenfunctions
  $\phi_1$ (long dashes) and $\phi_2$ (dashes and dots), and half the
  density $\rho/2 = |\phi_1|^2 + |\phi_2|^2$ (full thick line), obtained
  from the unperturbed Hamiltonian 
  $\hat{h}_0$, plotted as a function of $r$. Two perturbed potentials are
  also presented, namely $U_{\rm ion}-1.5\rho$ (short dashes) and
  $U_{\rm ion}+1.5\rho$ (dots).
\label{fig:model}
}
\end{center}
\end{figure}

We consider four electrons, two spin-up and two spin-down, in
one spatial dimension with the Hamiltonian 
\begin{subequations}
\label{eq:fullHam}
\begin{eqnarray}
  \hat{H}
  &=&
  \hat{h}_0+\hat{V}
  \quad, \quad 
  \hat{V} = \frac{g}{2}\sum_{i\neq j}\delta(r_i-r_j)
  \quad,
\label{eq:H}
\\
  \hat{h}_0
  &=&
  -\frac{\hat\Delta}{2}+U_\mathrm{ion}
  \quad,\quad
  U_\mathrm{ion} 
  =
    -\frac{e^2 q}{\sqrt{(r-r_0)^2+a^2}} 
    -
    \frac{e^2(Q-q)}{\sqrt{(r+r_0)^2+a^2}}
  \quad.
\end{eqnarray}
\end{subequations}
The electron-electron interaction $\hat{V}$ is taken schematically
as zero range. The external ionic potential is regularized at short distance.
Its parameters are chosen as $r_0=4$ $a_0$, $a=\sqrt{5}$ $a_0$, 
total charge $Q=2$ and charge in the right well $q=1.5$. The potential has two centers
around $\pm r_0$ and the charge $q$ creates a spatial asymmetry,
see Fig.~\ref{fig:model}.  The ratio $e^2/r_0$ sets
the natural energy unit.

In order to make the comparison of the different SIC schemes more
transparent, we study the interacting problem in a small basis of two
states. These basis states are generated from solving first the
unperturbed problem,
$\hat{h}_0 |\phi_{i\sigma}\rangle = \epsilon_i^{(0)}
|\phi_{i\sigma}\rangle$, 
where $\sigma\in\{\uparrow,\downarrow\}$ is the spin label and $i$
counts the level sequence which is generated for $\sigma$. Note that
$\hat{h}_0$ is hermitian and thus the $ |\phi_{i\sigma}\rangle$ are
orthonormal. 
The case is fully symmetrical in both spins. Thus we will
drop in the following the spin labels wherever this causes no
ambiguities. 
%
We take for further considerations the energetically lowest two
states, i.e., $i=1,2$.  The $\phi$ as well as the total density
in a given spin subspace, $\rho(r)/2 = |\phi_1(r)|^2 + |\phi_2(r)|^2$,
are plotted in Fig.~\ref{fig:model}. One can also see in this figure
the effect of an additional mean-field term $g\rho(r)$ for $g=-1.5$
and $g=1.5$. A negative coupling constant corresponds to an extra
attraction and the deepest well is deepened, while a positive coupling
constant gives a repulsion and tends to decrease the depth of the
wells.
The two-body interaction is taken into account at various levels of
mean-field approximation. The corresponding mean-field solutions
$\psi_{i\sigma}$ are expanded in the basis of the two occupied
unperturbed states, i.e.
$|\psi_{i\sigma}\rangle = c_{i1} |\phi_{1\sigma}\rangle + c_{i2}
|\phi_{2\sigma}\rangle$ 
(a more suitable parameterization will be introduced later).
This transformation 
redistributes components amongst occupied states and is
thus concentrating discussions particularly on the SIC.

\subsection{Energy expressions}

The various methods used here can all be formulated in terms of the
energy-density functional. The full HF case serves as a benchmark. The
HF energy as derived from the full Hamiltonian $\hat{H}$ (\ref{eq:H})
reads
\begin{equation}
  E^\mathrm{(HF)}
  =
  E_0
  +
  \frac{g}{4}\int \mathrm d r\,\rho(r)^2
  \quad,\quad
  E_0
  =
  2\sum_{i=1,2}\varepsilon_i
\label{eq:Eexch}
\end{equation}
where the factor 2 in $E_0$ stands for spin degeneracy and
$\rho(r)=2|\psi_1(r)|^2+2|\psi_2(r)|^2$. Here the $\psi$ denote the
eigenfunctions of the perturbed $\hat{H}$. Note that the zero-range
interaction $\hat{V}$ produces a purely density-dependent energy
already at the level of exact exchange. 
When only the Hartree contribution is taken into account, the energy
is now given by
\begin{equation}
  E^\mathrm{(Ha)}
  =
  E_0
  +
  \frac{g}{2}\int \mathrm d r\,\rho(r)^2
  \quad.
\end{equation}
This Hartree energy is deduced from the direct term of the interaction
only.  This raises the self-interaction problem. Augmenting that
Hartree energy by a self-interaction correction (SIC) reads
\begin{equation}
  E^\mathrm{(SIC)}
  =
  E_0
  +
  \frac{g}{2}\int \mathrm d r\,\rho(r)^2
  -
  \sum_{i,\sigma}
  \frac{g}{2}\int \mathrm dr\,{\rho_{i\sigma}(r)}^2
  \quad,\quad
  \rho_{i\sigma}(r)
  =
  |\psi_{i\sigma}(r)|^2
  \quad.
\label{eq:E-SIC}
\end{equation}

\section{The mean-field equations in  various approaches}
\label{sec:MF}

The mean-field equations are derived from a given energy expression
by variation with respect to the single-electron wavefunctions
$\psi_i$. This reads in general
\begin{equation}
  \hat{h}_i|\psi_i\rangle
  =
  \varepsilon_i|\psi_i\rangle
  \quad,\quad
  \hat{h}_i
  =
  \hat{h}_0
  +
  U^\mathrm{(mf)}_i
\label{eq:mfgeneral}
\end{equation}
where the self-consistent mean-field (mf) contribution
$U^\mathrm{(mf)}_i$ from the electron-electron interaction depends on
the actual level of approach.  The unperturbed part $\hat{h}_0$
remains the same in all approaches.  All further discussions
concentrate on the mean-field potential.  Note that the mean-field
Hamiltonian may depend on the state $i$ on which it acts. That will be
a major topic in the following.

\subsection{Hartree-Fock}
\label{sec:HF}

In the Hartree-Fock (HF) scheme, the Fock term automatically cancels
all self-interactions.  It is the most complete approach in the
variational space of Slater states and provides the reference
theory as we work at the level of exchange only. 
The mean-field potential $U^\mathrm{(mf)}$ then reads
\begin{equation}
\label{eq:U_HF}
  U^\mathrm{(HF)}[\rho] 
  = 
  \frac{g}{2}\rho(r) 
  \quad.
\end{equation}
It is hermitian and so becomes the total mean-field Hamiltonian 
$\hat{h}^{(\rm HF)} = \hat{h_0} + U^\mathrm{(HF)}$. 
For then, the solutions of Eq. (\ref{eq:mfgeneral}) are orthonormal.
The mean-field equations can also be expressed in terms of matrix
elements which reads, for the two-state model,
\begin{equation}
\label{eq:HF}
  \langle\psi_1|\hat{h}^\mathrm{(HF)}|\psi_2\rangle
  =
  0
  \quad,\quad
  \varepsilon^\mathrm{(HF)}_i 
  = \langle \psi_i | \hat{h}^\mathrm{(HF)} |\psi_i\rangle
  \quad.
\end{equation}

\subsection{Hartree}
\label{sec:Hartree}

The situation is very similar to HF, except for a different  factor in
front of the mean-field potential. We have now, for $U^\mathrm{(mf)}$,
\begin{equation}
\label{eq:U_Ha}
  U^\mathrm{(Ha)}[\rho] 
  = 
  {g}\rho(r) 
  \quad.
\end{equation}
The further handling proceeds as for HF in the previous subsection.
The off-diagonal elements have to fulfill
$\langle\psi_1|\hat{h}^\mathrm{(Ha)}|\psi_2\rangle = 0$
and diagonal ones define the Hartree single-particle energies
$\varepsilon^\mathrm{(Ha)}_i$.  Note that the Hartree scheme is here
considered as the analogue of the Local Density Approximation widely
used in DFT.

\subsection{SIC}
\label{sec:SIC}

Variation of the SIC  energy (\ref{eq:E-SIC}) yields for the
mean-field potential~:
\begin{equation}
\label{U_FSIC}
  U^\mathrm{SIC}_{i\sigma}
  = 
  g \rho(r) - g |\psi_{i\sigma}(r)|^2
  = 
  U^\mathrm{(Ha)}[\rho] 
  - 
  U^\mathrm{(Ha)}[|\psi_{i\sigma}(r)|^2]
  \quad.
\end{equation}
This mean-field potential now depends explicitely on the state on
which it acts. Orthonormality of the solution of the mean-field
equation (\ref{eq:mfgeneral}) is not guaranteed anymore. 
Several strategies are used to deal with that complication.
We will present three variants and compare them step by step.

\subsubsection{Explicit orthonormalization a posteriori}
\label{sec:OSIC}

A straightforward attack to the problem is to solve the eigenvalue
equations (\ref{eq:mfgeneral}) for each state separately and to apply
explicit orthonormalization {\it a posteriori}, e.g. by a Gram-Schmidt
procedure. This reads
\begin{subequations}
\label{eq:SICnaive}
\begin{eqnarray}
 &&  \hat{h}_1 |\psi_1\rangle = \varepsilon_1 |\psi_1\rangle
    \quad\Longrightarrow\quad
    \varepsilon_1\,,\,|\psi_1\rangle
    \quad,
\label{eq:SICnaive1}
\\  
 &&  \hat{h}_i |\psi_i\rangle = \varepsilon_i |\psi_i\rangle
     \quad\&\quad|\psi_i\rangle\perp \{ |\psi_1\rangle, \ldots ,
 |\psi_{i-1}\rangle \}
    \quad\Longrightarrow\quad
    \varepsilon_i\,,\,|\psi_i\rangle
    \quad.
\label{eq:SICnaivei}
\end{eqnarray}
\end{subequations}
The equations (\ref{eq:SICnaivei}) look seducing. However, one does
not solve the eigenvalue problem for $|\psi_i\rangle$ in full but only
in a restricted space where all states below have been projected out
by the orthogonalization.
We will denote that approach by the acronym ``OSIC''.  The price to be
paid for that simplification will be checked in our detailed example
below.

\subsubsection{Orthonormality by Lagrange multipliers}
\label{sec:LFSIC}

A more satisfactory scheme consists in using Lagrange multipliers to
impose the orthonormality constraint, $\langle \psi_j | \psi_j \rangle
= \delta_{ij}$, at the level of the variational
formulation following~\cite{Mes08}. We will refer to this
scheme by the acronym ``LSIC''. The system one has to solve is then
given by~:
\begin{subequations}

\label{eq:LSIC}
\begin{equation}
  \hat{h}_i |\psi_i\rangle 
  = 
  \sum_{j=1,2} \lambda_{ij}
  |\psi_j\rangle, 
  \qquad 
  i=1,2
\label{Schr_LFSIC}
\end{equation}
where $\hat{h}_i$ is defined in Eq.~(\ref{eq:mfgeneral}) and the Lagrange
multipliers are given by $\lambda_{ij} = \langle \psi_j | \hat{h}_i |
\psi_i \rangle$. Furthermore the orthonormality of the $\psi$ imposes
 a ``symmetry'' condition on the $\lambda_{ij}$~:
\begin{equation}
  \lambda_{ij} 
  = 
  (\lambda_{ji})^* 
  = \frac{1}{2} \langle \psi_i |\hat{h}_i + \hat{h}_j | \psi_j \rangle
  \quad.
\label{sym_LFSIC}
\end{equation}
\end{subequations}
That equation is more involved than Eq.~(\ref{eq:mfgeneral}), since it
is no longer an eigenvalue equation but they 
guarantee that the solutions $\psi_i$ will be orthogonal.
Furthermore, as Eq. (\ref{sym_LFSIC}) stems from a variational principle
on the energy in an effectively reduced space (orthonormal orbitals),
it is necessary fulfilled.

The constrained equations (\ref{eq:LSIC}) do not yield immediately
single-particle energies as the $\lambda_{ij}$ matrix is not diagonal. We 
thus define the
$\varepsilon_i^\mathrm{(LSIC)}$ as eigenvalues of the constraint
matrix $\lambda_{ji}$. Eq. (\ref{sym_LFSIC}) shows that $\lambda_{ji}$
is a hermitian matrix. Thus a diagonalization is possible and the
definition makes sense.

\subsection{Ignoring orthonormalization}

One could be very naive and ignore the orthonormality at all.  In our
model, that amounts simply to solve the eigenvalue problem with
$U_{1\sigma}$ associating the solution with lowest eigenvalue and with
$U_{2\sigma}$ associating the highest eigenvalue.  That naive SIC scheme
will be labeled by the acronym N-OSIC for ``non-orthonormalized SIC'',
in order to stress the loss of orthonormalization of the eigenstates.
We consider that, in principle prohibited, option for pedagogical
purposes. 

\section{Comparison between HF, Hartree and SIC}
\label{sec:compar}

\subsection{Numerical handling in the two-state model}
\label{sec:num}

The aim is to check the simultaneous fulfillment of the mean-field
equations (\ref{eq:mfgeneral}) for each state $i=1,2$ in the two-state
model together with orthonormality of the corresponding solutions $\psi_i$.
 We restrict the solutions to the configuration space spanned
by the two energetically lowest eigenstates $\phi_1$ and $\phi_2$ of
the unperturbed problem, see section \ref{sec:model}.  The solutions
of the interacting mean-field equations will thus be expanded as
\begin{equation}
  \left( \begin{array}{c} \psi_1 \\  \psi_2 \end{array} \right) 
  =
  \left( \begin{array}{rcr} 
    \cos\theta_1 && -\sin\theta_1 \\ 
    \sin\theta_2 &\ & \cos\theta_2 
  \end{array} \right) 
  \left( \begin{array}{r} \phi_1 \\  \phi_2 \end{array} \right) 
  \quad,
\label{eq:transf}
\end{equation}
the same way for both spins. That transformation maintains
normalization of the $\psi$. It becomes a unitary transformation if
$\theta_2=\theta_1$ and then also keeps orthogonality between the
$\psi$. However we {\it a priori} start with $\theta_2\neq\theta_1$.

Solution  of mean-field equations means that the variance of the
corresponding mean-field Hamiltonian becomes zero. Thus we consider
as a  global measure of convergence the squared deviations from the goal,
\begin{equation}
\label{eq:chi2}
  \chi^2
  =
  \frac{\sigma^2} {\langle h \rangle^2}
  +
  \sin^2(\theta_2-\theta_1)
  \quad,\quad
  \sigma^2 = \sum_i \Delta h^2_i
\end{equation}
where the mean field variances for HF, Hartree, N-OSIC and OSIC read
\begin{subequations}
\label{variance_h}
\begin{equation}
  \Delta h^2_i
  =
  \sum_j\Big|\langle \phi_j |\hat{h}_i - \varepsilon_i| \psi_i \rangle\Big|^2 
  \quad,\quad
  \varepsilon_i
  =
  \langle \psi_i | \hat{h}_i | \psi_i \rangle   
  \quad,
\label{variance_FSIC}
\end{equation}
while in the case of LSIC, these expressions are more involved, due
to the Lagrange multipliers $\lambda_{ik}$,
\begin{equation}
  \Delta h^2_i 
  = 
  \sum_{j} \left|\left\langle \phi_j \Big|
    \hat{h}_i 
    - 
 \sum_k \lambda_{ik}
  \Big| \psi_i\right\rangle\right|^2
  \quad,\quad
   \lambda_{ik}
   =
   \frac{1}{2} \langle\psi_i|\hat{h}_i+\hat{h}_k|\psi_k\rangle \quad .
\label{variance_LFSIC}
\end{equation}
\end{subequations}
The mean value of $\hat{h}$  is defined as $\langle h
\rangle = \frac{1}{2} \sum_i  \langle \psi_i | \hat{h}_i |
\psi_i \rangle$. 
The solution is found by searching the global  minimum of
$\chi^2$ in the space of the angles $\theta_1$ and $\theta_2$.
The natural result  $\theta_1=\theta_2$ emerges immediately for
HF, Hartree and LSIC. The case remains open for N-OSIC and OSIC.

\subsection{Result and discussion}

Four coupling constants $g$ have been tested~: $-1.5$, $-0.5$, 0.5 and
1.5. For all 
cases, the HF single particle energies correspond to bound states and
the shifts in energy are rather small, consistently with the fact that our
zero-range interaction potential is a perturbation of $\hat{h}_0$, 
as we work in the basis of the  eigenstates of $\hat{h}_0$. For $g>1.5$, the
repulsion starts to be too strong and the eigenstates are less and
less bound, or even not bound anymore.

\begin{figure}[htbp]
\begin{center}
\epsfig{file=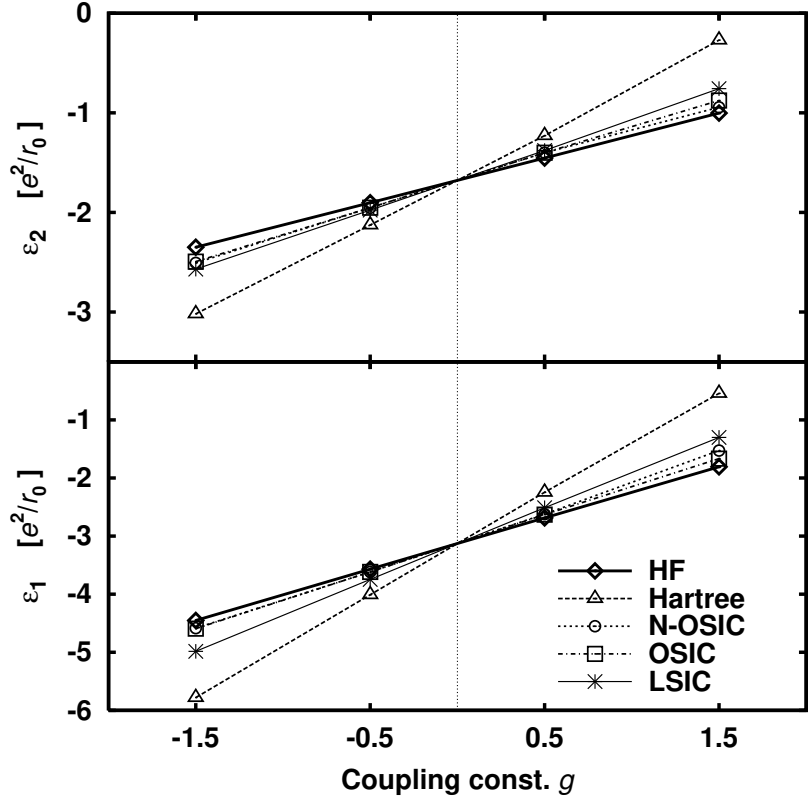,width=0.8\linewidth}
\caption{Single particle energies of the eigenfunctions obtained
  from the perturbed Hamiltonian within different approaches~: HF
  (circles), Hartree (triangles), SIC without orthogonalization (N-OSIC,
  circles), SIC with orthogonalization (OSIC, squares), and SIC with
  Lagrange multipliers (LSIC, crosses).
\label{fig:energy}
}
\end{center}
\end{figure}

In Fig.~\ref{fig:energy}, we present the two eigen-energies
$\varepsilon_1$ and $\varepsilon_2$ for the various schemes. 
Remind that the HF values represent our benchmark calculations.
As expected, the Hartree calculation, which does not account for any SIC, 
gives values that are far from the HF energies. For negative coupling
constants, the Hartree states are too bound, while for positive values, they
are less bound. This is not a surprise, since comparing
Eqs.~(\ref{eq:U_HF}) and (\ref{eq:U_Ha}), $U^{\rm (Ha)}$ is twice
larger than $U^{\rm (HF)}$.

When SIC is included, the eigen-energies are very close to the HF
results. Surprisingly SIC without imposing the orthogonality of the
$\psi$ (N-OSIC) works remarkably well, and actually the corresponding
energies are indistinguishable from those of SIC with orthogonality.

\begin{figure}[htbp]
\begin{center}
\epsfig{file=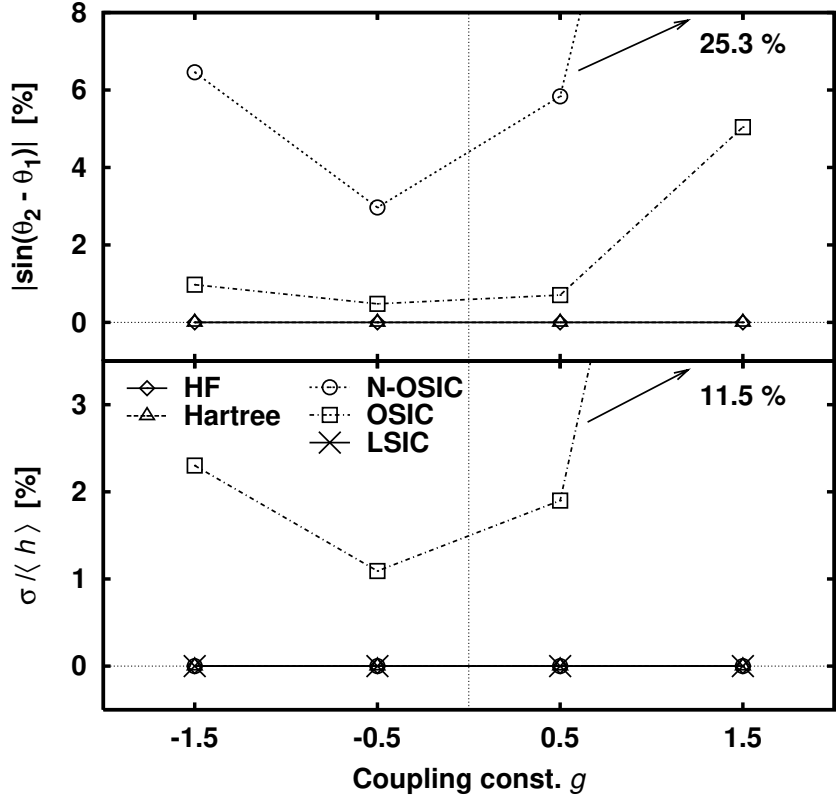,width=0.8\linewidth}
\caption{Ratio of standard deviation over mean value of the
  Hamiltonian, $\sigma/\langle h \rangle$, and violation of 
  orthogonality of the $\psi$, $|\sin(\theta_2 - \theta_1)|$,
  as a function of the coupling constant $g$~: HF (diamonds), Hartree
  (triangles), SIC without orthogonalization (N-OSIC, circles), SIC
  with imposed orthogonality (OSIC, 
  squares), and SIC with Lagrange multipliers (LSIC, crosses). In the
  upper panel, the orthogonality violation has no meaning for LSIC. In
  both panels, values for $g=1.5$ are out of range, as indicated.
\label{fig:indicators}
}
\end{center}
\end{figure}

However our concern is precisely the conservation of
orthogonality. Following that aim, we have plotted in 
Fig.~\ref{fig:indicators} two indicators of the resolution precision
for each scheme~: the ratio of the Hamiltonian variance over its mean
value, $\sigma/\langle h \rangle$, see 
Eq.~(\ref{eq:chi2}), for all SIC schemes, and the orthogonality
violation of the $\psi$, $|\sin(\theta_2 - \theta_1)|$, only in the case
of HF, Hartree, SIC, N-OSIC and OSIC. Note that such an indicator should
not exist neither for OSIC nor for LSIC, since the orthogonality of
the $\psi$ is explicitely taken into account in both schemes. And
indeed $|\sin(\theta_2 - \theta_1)|$ vanishes for LSIC as it should,
for the orthonormalization of the $\psi$ is imposed in the variational
derivation of the total energy. However in OSIC, as is discussed in
Sec.~\ref{sec:OSIC}, the brute force orthogonalization restricts the
space of the $\psi$ and a vanishing minimum of the $\chi^2$,
Eq.~(\ref{eq:chi2}), may not exist. And this is indeed the case as we
shall see below.

As expected, HF and Hartree, which are hermitian, give perfect indicators
in the sense that they are equal to zero~: $\sigma/\langle h \rangle$
is about $10^{-16}$, while the orthogonality of the $\psi$ is verified
within an error less than $10^{-15}$. In LSIC, we also obtain the same
order of magnitude for $\sigma/\langle h \rangle$.

Now let us focus on the various SIC we have tested. In N-OSIC, 
Eqs.~(\ref{eq:SICnaive}), the standard deviation of $h$ is one order
of magnitude higher than in LSIC but remains very small
(less than $10^{-14}$). However, since no orthogonality has been taken
into account, $|\sin(\theta_2 - \theta_1)|$ does not vanish, as
expected~: it ranges from 3~\% up to 25~\%.
When one minimizes at the same time the standard deviation of $h$ and
the orthogonality condition (OSIC scheme), one of course obtains a better
indicator for the orthogonality of the $\psi$, as is visible in
Fig.~\ref{fig:indicators} when comparing circles (N-OSIC) with squares
(OSIC). In the latter case, the orthogonality is violated by
0.4--5~\%. But note that  $\sigma/\langle h \rangle$ does not vanish as
well~: For small couplings, it varies between 1~\% and 2~\%, 
while it goes up to 11.5\% in the worst case. This means that, in this
example, it is impossible to solve in a satisfactory way 
Schr\"odinger equations with orthogonal eigenstates. And
actually, the violation is shared among the orthogonality and 
the variance of $h$. 

\begin{figure}[htbp]
\begin{center}
\epsfig{file=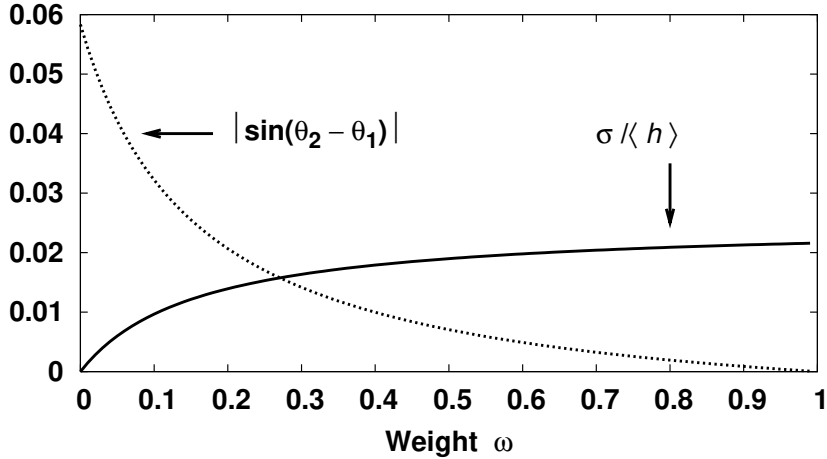,width=0.8\linewidth}
\caption{Ratio of standard deviation over mean value of the
  Hamiltonian, $\sigma/ \langle h \rangle$, (full line) and
  orthogonality  of the $\psi$, $|\sin(\theta_2 - 
  \theta_1)|$, (dotted line) as a function of the
  relative weight $\omega$ in the minimization of $f(\omega) =
  (1-\omega) \sigma^2/\langle h \rangle^2 + \omega \sin^2(\theta_2 -
  \theta_1)$. 
\label{fig:weight}
}
\end{center}
\end{figure}

To end this section, we come back to the OSIC scheme. As stated just
above, the violation of the vanishing of both orthogonality and
variance of $h$ is shared among these two constraints. One can
actually give 
more weight to one condition with respect to the other one, for
instance to the Schr\"odinger equations. In Fig.~\ref{fig:weight}, we
have plotted the results of the minimization of
\begin{equation}
f(\omega) = (1-\omega) \frac{\sigma^2}{\langle h \rangle^2} +
\omega\sin^2(\theta_2 - \theta_1)\quad , 
\end{equation}
where $\omega$ is a weight varying between 0 and 1, for a coupling
constant of 0.5. 
We note that indeed, whatever the weight, both constraints are violated
by the same order of magnitude. This means once again that it is
impossible to meet both conditions at the same time in the OSIC
scheme.



\section{Conclusion}
\label{sec:ccl}

We have investigated the problem of orthonormality of the occupied
single-particle states in mean-field equations with self-interaction
correction (SIC). To that end, we have used a simple two-state model
with zero-range interaction which concentrates fully on the share
amongst the occupied states through SIC. The full Hartree-Fock (HF) case
serves as a benchmark. The HF Hamiltonian is hermitian and we have
energy diagonality together with orthonormality of the states.  The
same holds for the Hartree approach which, however, is plagued by the
self-interaction error. SIC produces a state-dependent Hamiltonian and
orthonormality becomes an issue. Naively ignoring that, yields
deceivingly good results for the energies but awfully wrong
wavefunctions, with disastrous consequences for other observables.
Taking care by explicit Gram-Schmidt orthonormalization during the
mean-field solution provides acceptable results leaving, however,
non-negligible variances in the energy. The error can remain small 
depending on the case. The consistent scheme dealing with a matrix of
Lagrange multipliers is a bit involved and requires a separate step to
defined single particle energy. But it is the only scheme delivering
consistent and satisfying SIC results.



Note that, in static calculations, the orthonormalization may be a
marginal problem in many cases as observables are found to stay very
close to the HF ones.  However it will build up to large errors in
time-dependent cases and could even lead to divergencies.

\end{document}